\documentclass[reprint, superscriptaddress, amsmath, amssymb, aps, prb]{revtex4-2}

\usepackage{graphicx}
\usepackage{amsmath}
\usepackage{amssymb}
\usepackage{subcaption}
\usepackage{bm}
\usepackage{xcolor}

\usepackage[mathlines]{lineno}

\usepackage[colorlinks=true, linkcolor=blue,urlcolor=blue,citecolor=blue]{hyperref}
\usepackage{cleveref}

\usepackage[ignoreunlbld,norefs,nocites]{refcheck}

\usepackage{hyperref}
\usepackage[justification=justified,singlelinecheck=false]{caption}

\usepackage{caption}
\usepackage{caption}
\begin{document}
	
	\preprint{APS}

	\title{Many body localization in Disordered
		One-Dimensional Fermi-Hubbard Model
	}
	
	\author{Harmanjeet Kaur}
	\affiliation{%
		Computational Quantum Many-Body Physics Lab, Department of Physics, Dr.\ B.\ R.\ Ambedkar National Institute of Technology, Jalandhar, Punjab - 144008, India}%

	\author{Vinod Ashokan}
	\email{ashokanv@nitj.ac.in}
	\affiliation{Computational Quantum Many-Body Physics Lab, Department of Physics, Dr.\ B.\ R.\ Ambedkar National Institute of Technology, Jalandhar, Punjab - 144008, India}

\begin{abstract}
We investigate the non-equilibrium dynamics of the disordered one-dimensional Fermi-Hubbard model with a focus on many-body localization. The system is initialized in a charge-density-wave-state, and its time evolution is analyzed through sublattice imbalance (spin and charge), and bipartite entanglement entropy. A clear crossover from ergodic to non-ergodic behavior is observed with increasing disorder strength. In the weak disorder regime, rapid decay of imbalance and the fast growth of entanglement indicate efficient thermalization. In contrast, a strong disorder leads to persistent imbalance and slow dynamics, signaling the breakdown of ergodicity. The charge and spin sectors exhibit distinct relaxation behavior, providing evidence for partial decoupling between these degrees of freedom. Furthermore, in the interacting regime, the entanglement entropy shows slow logarithmic growth, reflecting the dephasing-driven dynamics characteristic of the many-body localized phase. These results highlight the interplay between disorder and interactions in determining the dynamical properties of the system and establish robust signatures of many-body localization in the Fermi-Hubbard model. 
\end{abstract}

	\maketitle

\section{Introduction}
The nonequilibrium dynamics of isolated interacting quantum systems are typically governed by the eigenstate thermalization hypothesis (ETH) \cite{PhysRevA.43.2046,PhysRevLett.116.247204}, which predicts relaxation toward thermal equilibrium. However, this paradigm breaks down in the many-body localized (MBL) phase \cite{Rispoli2019}, where sufficiently strong disorder in the presence of interactions inhibits thermalization and gives rise to non-ergodic behavior \cite{Nandkishore2015,Sierant2022,Mondaini2015}. As the consequence, localized systems retain memory of their initial conditions and fail to approach thermal equilibrium even at long time. Owing to these unusual non-equilibrium properties, MBL has remained an active and rapidly developing field of research in condensed matter physics \cite{BASKO20061126,Oganesyan2007,Huse2014,Abanin2017,Ovadia2015}. Most theoretical studies of MBL have focused on disordered spinless fermion systems \cite{Kelly2020}, whereas ultracold-atom experiments are more accurately described by interacting disordered Hubbard models containing spinful fermions \cite{Ponte2017}. Due to the presence of both charge and spin degrees of freedom, the Hubbard model exhibits considerably richer non-equilibrium dynamics than its spinless counterpart. Previous investigations have shown that, under strong disorder, the initially prepared density imbalance can persist for long times, signaling non-ergodic behavior in agreement with experiment observations \cite{Bordia2016}.

An important feature of the disordered Hubbard model is the distinct response of charge and spin sectors to disorder and interactions. For finite onsite interaction $U>0$, a strong potential disorder suppresses charge transport , causing the charge-density-wave observables and local charge correlations to retain memory of the initial configuration and fail to thermalize. In contrast, spin observables, such as spin imbalance and local spin correlations, continue to evolve and decay with time. The coexistence of these two dynamical behaviors is also reflected in the evolution of the entanglement entropy, which incorporates contributions from both charge and spin excitations. 
The spin sector initially shows a near-linear growth, followed by a crossover to logarithmic behavior at later times, while the charge sector exhibits predominantly logarithmic dynamics over the entire evolution. The resulting slow growth of the entanglement entropy provides a characteristic signature of constrained thermalization and localization effects in interacting disordered systems.
These results indicate that strongly disordered Hubbard systems may not realize a full MBL phase characterized by complete set of local conserved quantities~\cite{Nandkishore2015}. Rather, they point toward a regime of partial non-ergodicity accompanied by an effective dynamical separation between charge and spin degrees of freedom. Furthermore, complete localization of the spin sector can emerge when the symmetry between spin-up and spin-down fermions is broken, for example through the introduction of spin-dependent disorder. In this work, we investigate the non-equilibrium dynamics of the one-dimensional disordered Fermi-Hubbard model following a global quantum quench. Employing exact diagonalization, we study the time evolution of charge and spin imbalance observables together with the corresponding entanglement entropy over a broad range of disorder strengths and interaction parameters.  Our analysis focuses on the distinct relaxation dynamics of the charge and spin sectors and their relation to localization phenomena in interacting disordered systems.These findings provide further evidence for partial localization and highlight the important role of spin-charge separation in determining the non-equilibrium dynamics of interacting disordered fermionic systems.

The rest of the paper is structured as follows. In Sec \ref{sec:model}, we introduce the disordered Fermi-Hubbard Hamiltonian and discuss the model parameters. In \ref{sec:observables}, we define the non-equilibrium observables used to characterize localization, including the charge imbalance, spin imbalance, and bipartite entanglement entropy. Section \ref{sec: results} and discussion of the charge, spin, and entanglement dynamics. Finally, \ref{sec:conclusion} summerizes the main conclusions and outlook.

\section{ Theoretical Model}
\label{sec:model}
\subsection{Fermi-Hubbard Hamiltonian}
To study the Many-Body Localization phenomenon theoretically and experimentally, one of the most important models used is the Fermi–Hubbard model. The Fermi-Hubbard model with quenched random disorder which has the following Hamiltonian:

\begin{equation}
	\begin{aligned}
		H = &-J \sum_{i=0,\sigma}^{L-2} \left(c^\dagger_{i\sigma} c_{i+1,\sigma} - c_{i\sigma} c^\dagger_{i+1,\sigma} \right) \\
		&+U  \sum_{i=0}^{L-1}n_{i\uparrow}n_{i\downarrow}
		+ \sum_{i=0,\sigma}^{L-1}V_in_{i\sigma}
	\end{aligned}
\end{equation}
where the first term is the kinetic energy term and contains $J$, which is the tunneling matrix element between two adjacent lattice sites. The operators $c_{i\sigma}^{\dagger}$ and $c_{i\sigma}$ are fermionic creation and annihilation operators for a particle in spin state $\sigma$ at the lattice site $i$ and $i+1$. The second term describes the interaction energy in the system and $U$ is the on-site interaction. The occupation number of site $i$ is given by $n_{i\sigma}$. The last term is the disorder term where $V$ describes the disorder potential and a random on-site disorder potential \cite{Weinberg2019}. The final term introduces quenched disorder through the random on-site potential $V_i$ , where the disorder values are chosen from a uniform distribution within the range $
V_i \in[-W,W]
$. Here, \(W\) characterizes the strength of the disorder and serves as the primary control parameter of the system. For weak disorder, the system remains ergodic and exhibits thermalization as a result of particle transport and interactions. As the strength of disorder increases, the motion of particles becomes increasingly suppressed, eventually leading to localization and the emergence of the many-body localized phase \cite{Abanin2019}.
\section{ Quantum Dynamics and Observables}
\label{sec:observables}
In the non-equilibrium dynamics, the system is initialized in a charge density wave (CDW) state, where fermions occupy alternating lattice sites. Such an initial configuration provides a direct probe of transport and relaxation processes in the presence of interactions and disorder. During time evolution, the decay of the initial density modulation reflects ergodic behavior and thermalization, whereas its long-time persistence signals suppressed the transport and localization effects \cite{Schreiber2015}.

\subsection{Sublattice Imbalance}

To characterize the non-equilibrium dynamics and the persistence of the initial density modulation, we consider the sublattice imbalance, defined as the difference in occupation between even and odd lattice sites. This observable is particularly well-suited for systems initialized in a CDW state \cite{Schreiber2015}.
The sublattice-resolved particle number is defined as 
\begin{equation}
    N_{even}(t) = \sum_{i\in{even} }\langle n_i(t) \rangle  , \quad  N_{odd}(t) = \sum_{i \in odd} \langle n_i (t) \rangle 
\end{equation}
where $ n_i = n_{i \uparrow} +n_{i \downarrow} $ is the local particle density.

The sublattice imbalance is then given by
\begin{equation}
    I(t) = \frac{N_{\text{even}}(t) - N_{\text{odd}}(t)}{N_{\text{even}}(t) + N_{\text{odd}}(t)}
\end{equation}
The imbalance can be written more compactly as $
     I(t) = \frac{1}{N}\sum_i(-1)^i\langle n_i(t)\rangle.$
This quantity provides a direct measure of how the initial density modulation evolves over time. In an ergodic phase, particle transport leads to a uniform redistribution of the density, causing the imbalance to decay to zero. In contrast, if the system is in a localized regime, the imbalance retains a finite value even at long times, indicating that the system preserves its initial memory \cite{Bordia2017, Luschen2017}.

\subsection{Spin-Charge Imbalance}
To characterize the non-equilibrium dynamics of the disordered one-dimensional Fermi-Hubbard model, we study the time evolution of charge and spin imbalance observables following a quantum quench from highly ordered initial product states. The system is initialized either in a charge-density-wave (CDW) configuration or in a spin-density-wave (SDW) configuration, allowing independent investigation of the charge and spin sectors \cite{PhysRevB.107.L180201}.
The charge dynamics are probed using the normalized charge imbalance $I_c(t) = 1/N\sum_i(n_{i\uparrow}(t) + n_{i\downarrow}(t))$ where $n_{i\sigma} = c^\dagger_{i,\sigma}c_{i,\sigma}$ denotes the local density operator for spin $\sigma = \uparrow,\downarrow$. The simulations are initialized in the CDW product state 
$|\psi_c(0)\rangle = |\uparrow\downarrow \;, 0 \;, \uparrow\downarrow ,\; 0 \; ,\cdots \rangle$, corresponding to doublons occupying alternating lattice sites. The charge imbalance measures the persistence of the initial staggered density modulation during the time evolution and therefore provides a direct probe of charge transport and localization effects in the interacting disordered system \cite{BASKO20061126}. To investigate spin dynamics, we compute the normalized spin imbalance $I_s(t)=\frac{1}{N}\sum_i (-1)^i \left\langle n_{i\uparrow}(t)-n_{i\downarrow}(t)\right\rangle$ starting from the spin-density-wave initial state $
|\psi_s(0)\rangle = |\uparrow \;, \downarrow \;, \uparrow \;, \downarrow \;, \cdots \rangle
$ which contains alternating spin orientations on neighboring sites. Unlike the charge sector, the spin imbalance probes the relaxation of staggered spin order and provides direct information about spin transport in the presence of disorder and interactions \cite{Anderson1958, Zakrzewski2018, Kucsko2018}.

\subsection{Entanglement Entropy Dynamics}
The growth of entanglement entropy provides an important characterization of the dynamical behavior in Many-body Localization systems. To quantify this, the one-dimensional chain is bipartitioned into two equal halves at its center, each of length L/2, defining subsystems \textbf{A} and \textbf{B}. At any time t, the state of the entire system is denoted by $|\psi(t)\rangle$, and the corresponding density matrix is $\rho(t) = |\psi(t)\rangle \langle\psi(t)|$.
The reduced density matrix of subsystem \textbf{A} is obtained by tracing out the degrees of freedom of subsystem \textbf{B}:
\[
\rho_A(t) = \mathrm{Tr}_B[|\psi(t)\rangle \langle\psi(t)|].
\]
The entanglement entropy is then defined as the von Neumann entropy associated with $\rho_A(t)$\cite{Bardarson2012}:
\begin{equation}
S(t) = -\mathrm{Tr}(\rho_{A}\mathrm{ln}\rho_A) 
\end{equation}

In numerical simulations based on matrix product state methods, the wavefunction is naturally expressed in its Schmidt decomposition across the bipartition. In this representation, the reduced density matrix becomes diagonal with eigenvalues given by Schmidt coefficients $\lambda_i$, allowing an efficient evaluation of the entanglement entropy as $S(t) = -\sum_{i}\lambda_{i}\mathrm{ln}\lambda_i$,
where $\lambda_i$ represents the weights associated with the Schmidt links connecting the two halves of the chain \cite{Khemani2017, Lukin2019, Serbyn20131}.

The time dependence of $S(t)$ provides crucial insight into the nature of quantum dynamics. In ergodic systems, entanglement entropy typically grows rapidly due to the spreading of correlations and transport processes. In contrast, in strongly disordered interacting systems described by the Fermi-Hubbard model, disorder suppresses particle transport, while interactions induce slow dephasing. As a result, the system may enter many-body localized phase, where the entanglement entropy exhibits a characteristic logarithmic growth in time, $S(t)\sim \mathrm{log}t$. This slow growth originates from exponentially decaying interactions between localized degrees of freedom, which generate entanglement through dephasing rather than particle transport \cite{Serbyn20132}. Thus, the observation of logarithmic entanglement growth serves as a key signature of localization in the disordered Fermi–Hubbard model \cite{Serbyn2015}.

\section{Numerical Calculations and Results}
\label{sec: results}

\subsection{Spin-Charge Separation Dynamics}
\begin{figure}[htbp]
	\centering
	
	\begin{subfigure}[b]{1\columnwidth}
		\centering
		\includegraphics[width=\linewidth]{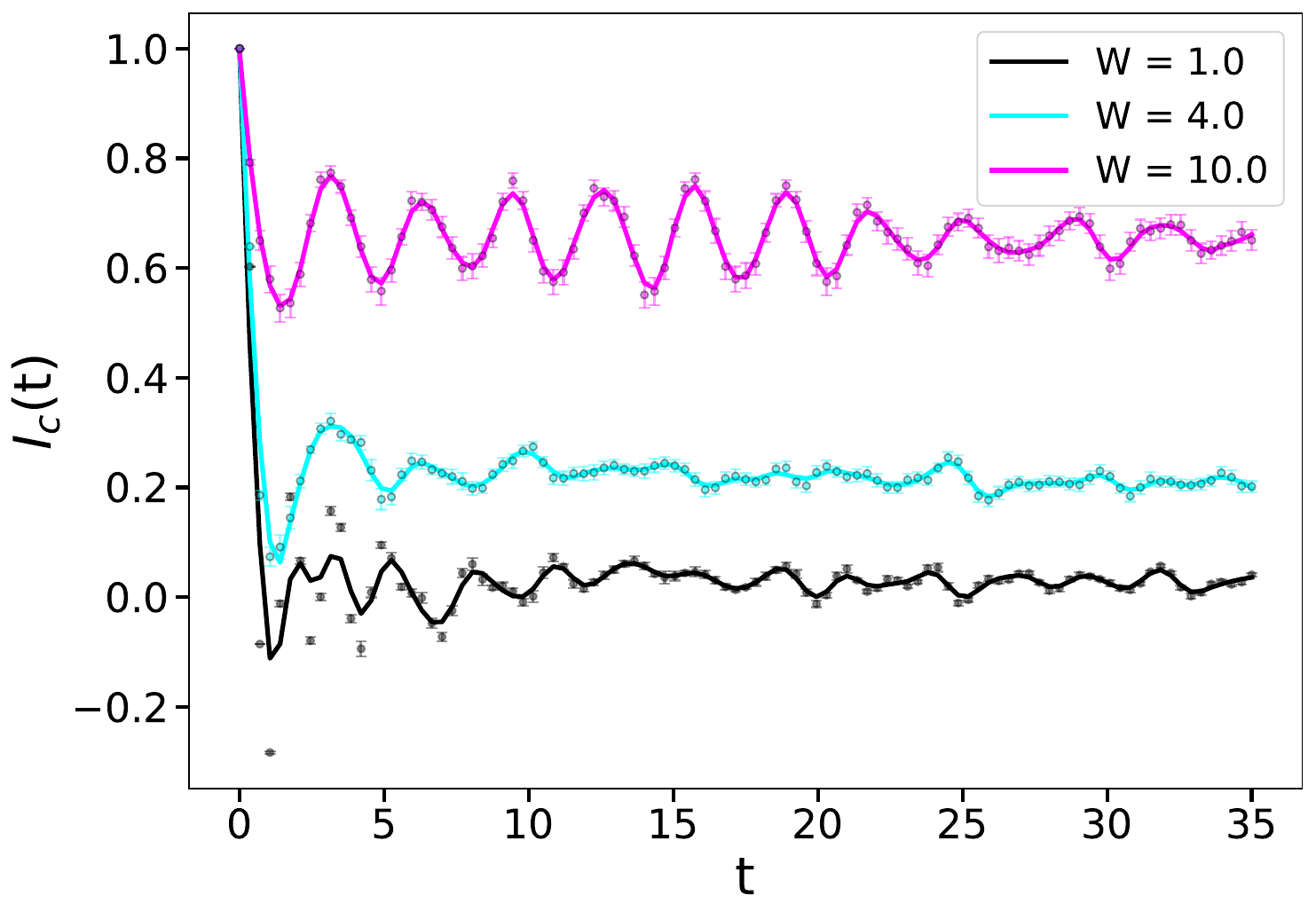}
		\caption{}
		\label{fig:Charge_imbalance}
	\end{subfigure}
	
	\vspace{0.3cm}
	
	\begin{subfigure}[b]{1\columnwidth}
		\centering
		\includegraphics[width=\linewidth]{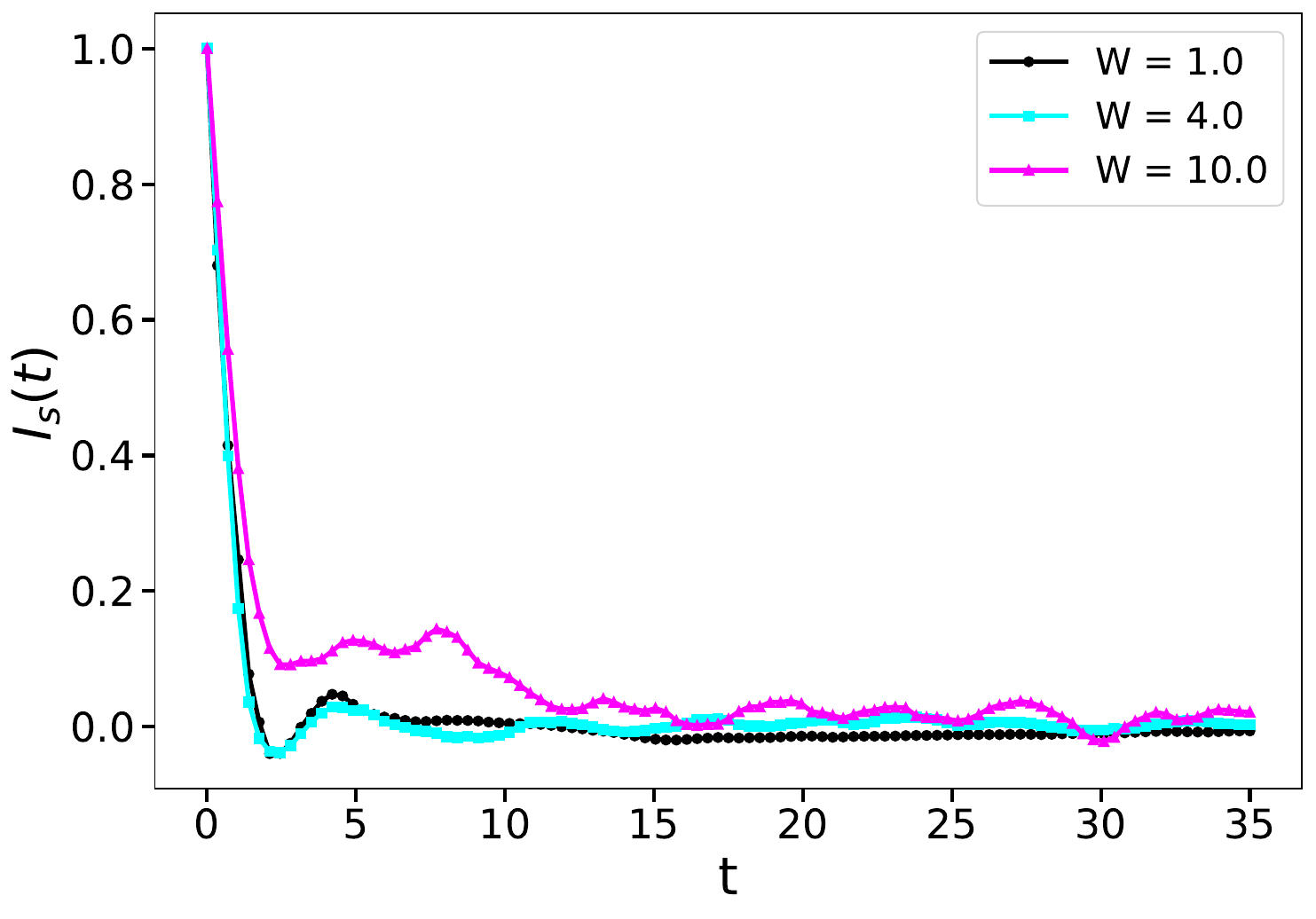}
		\caption{}
		\label{Spin_imbalance}
	\end{subfigure}
	
	\caption{Time evolution of (a) Charge Imbalance $I_C (t)$ and (b) Spin Imbalance $I_S (t)$ for system sizes $L = 12$ at $J=1$, $U = 5$ and disorder strengths $W = 1.0, 4.0, 10.0$.}
	\label{fig:Sublattice_imbalance}
\end{figure}

Figure~\ref{fig:Sublattice_imbalance} shows the disorder-average time evolution of both the charge imbalance $I_C(t)$ and spin imbalance $ I_S (t)$ for the one-dimensional disordered Fermi-Hubbard model with system size $ L = 12$ at different disorder strengths. The charge sector is initiated in a CDW with doublons occupying alternating lattice sites \cite{Weinberg2019}, whereas the spin sector is initialized in a SDW configuration with alternating spin occupations.
For weak disorder $W = 1.0$ as in Fig.~\ref{fig:Sublattice_imbalance} (a) and (b), both charge and spin imbalances exhibit a rapid decay toward zero, indicating efficient relaxation and ergodic dynamics. In this regime, the disorder potential is insufficient to significantly hinder particle motion, and the kinetic hopping term dominates the dynamics. Consequently, initial density-wave order melts quickly and the system approaches a spatially homogeneous thermal state at long times \cite{Abanin2019}.

As the disorder strength  increases that is $W = 4.0$ and $W=10.0$, as shown in Fig.~\ref{fig:Sublattice_imbalance} (a) the relaxation dynamics in charge sectors becomes progressively slower and develops a pronounced finite long-time remanant. At intermediate disorder strengths, residual hopping processes still permit partial relaxation of the initial CDW  patterns. However, in the strong disorder regime, the random on-site potential strongly suppresses transport, significantly reducing the propagation of charge  excitations. As a result, the charge imbalances $I_C (t)$ retain a substantial fraction of their initial values over accessible simulation timescales. 

The persistence of a finite long-time imbalance therefore reflects the gradual breakdown of ergodic transport regime to a strong localized phase characterized by slow relaxation dynamics and long-lived memory of the initial state. Furthermore, the comparatively slower relaxation observed at stronger disorder indicates enhanced localized effects arising from the interplay between interactions and quanched disorder, consistent with the established phenomenology of interacting many-body localized systems \cite{BASKO20061126}. In contrast as in Fig. \ref{fig:Sublattice_imbalance}  (b), for the disorder strength $W = 4.0$ and $W=10.0$ the spin sector $ I_S (t)$ remains delocalized and retains ergodic characteristics. 

\begin{figure}[htbp]
    \centering
    \includegraphics[width=\linewidth]{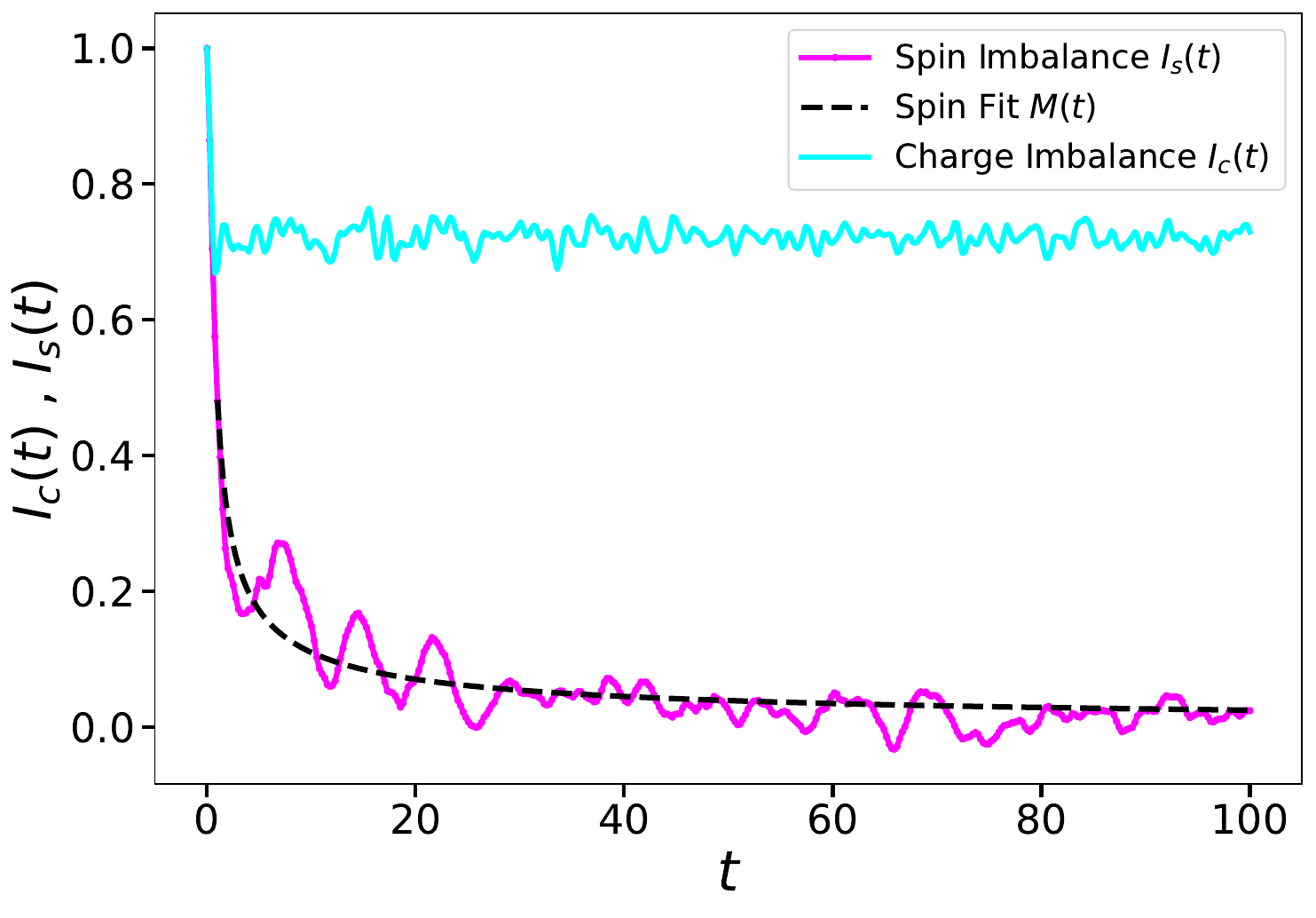}
    \centering
    \caption{The spin and charge imbalance of L = 12 at J=1, U = 4 and disorder strength W = 14.0 with fitted algebraic form M(t).}
    \label{spin_charge_L12}
   \end{figure}

The numerical results obtained from exact diagonalization reveal a pronounced dynamical separation between the charge and spin sectors in the disordered one dimensional  Fermi-
Hubbard model. The distinction is directly reflected in the time evolution of the charge and spin observables as shown in Fig. \ref{spin_charge_L12} for the spin and charge imbalance of $L = 12$ at $J=1$, $U = 4 $ and disorder strength with
$W = 14.0$. While charge imbalance probes the persistence of initial density modulation, the spin imbalance characterizes the relaxation of staggered spin order and therefore provides direct information about spin transport in the presence of disorder and interaction \cite{Anderson1958, Zakrzewski2018}.

   The charge imbalance exhibits a rapid initial reduction followed by saturation to a finite long-time value, reflecting strong suppression of charge transport in the presence of disorder. In contrast, the spin imbalance does not display clear saturation within the accessible simulation timescales and instead continues to decay gradually throughout the evolution. The long-time behavior of the spin sector is accurately captured by the algebraic form $ M(t)\sim At^{-\alpha}$ with the fitted exponent $(A \approx 0.4829 ,\alpha \approx 0.642)$. The observed power-law relaxation indicates slow non-equilibrium spin dynamics and suggests that the spin degrees of freedom remain partially delocalized even when the charge transport is strongly inhibited. Such algebraic decay is generally associated with anomalously slow thermalization in interacting disordered systems.

Our findings are consistent with experimental observations from cold-atom realizations of the quarter-filled one-dimensional Hubbard model, where a strong quasiperiodic potential leads to the emergence of a long-lived nonergodic charge imbalance. Extending these observations, we demonstrate that a random potential of comparable strength likewise renders the charge sector nonergodic. In contrast, spin correlations gradually vanish over time, indicating delocalized spin dynamics and the absence of localization in t  he spin sector. However, complete localization of both charge and spin degrees of freedom can be recovered by introducing spin-dependent disorder.

\subsection{Entanglement entropy}

\begin{figure}[htbp]
	\centering
	
	\begin{subfigure}[b]{1\columnwidth}
		\centering
		\includegraphics[width=\linewidth]{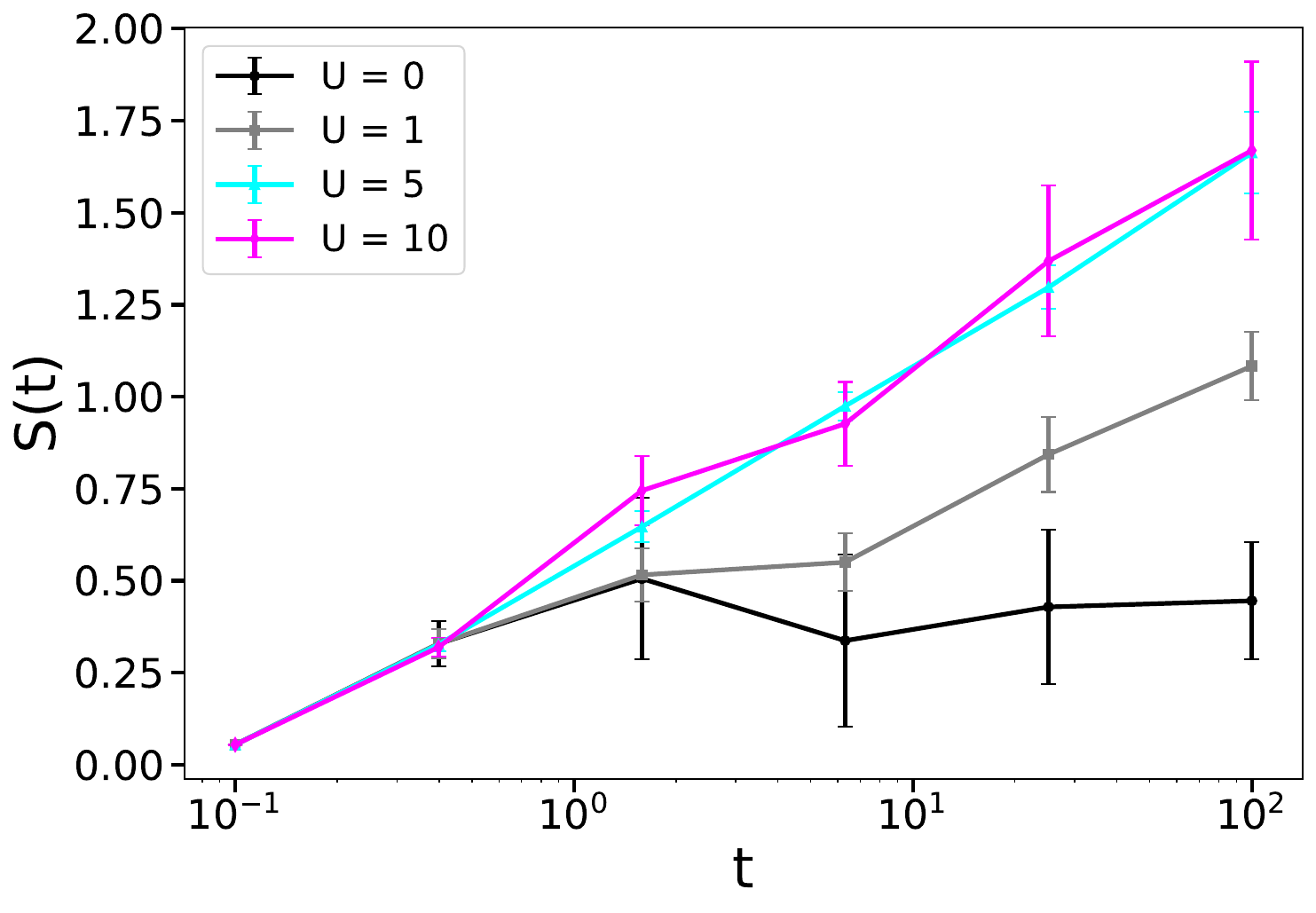}
		\caption{}
		\label{Fig: entanglement entropy}
	\end{subfigure}
	
	\vspace{0.3cm}
	
	\begin{subfigure}[b]{1\columnwidth}
		\centering
		\includegraphics[width=\linewidth]{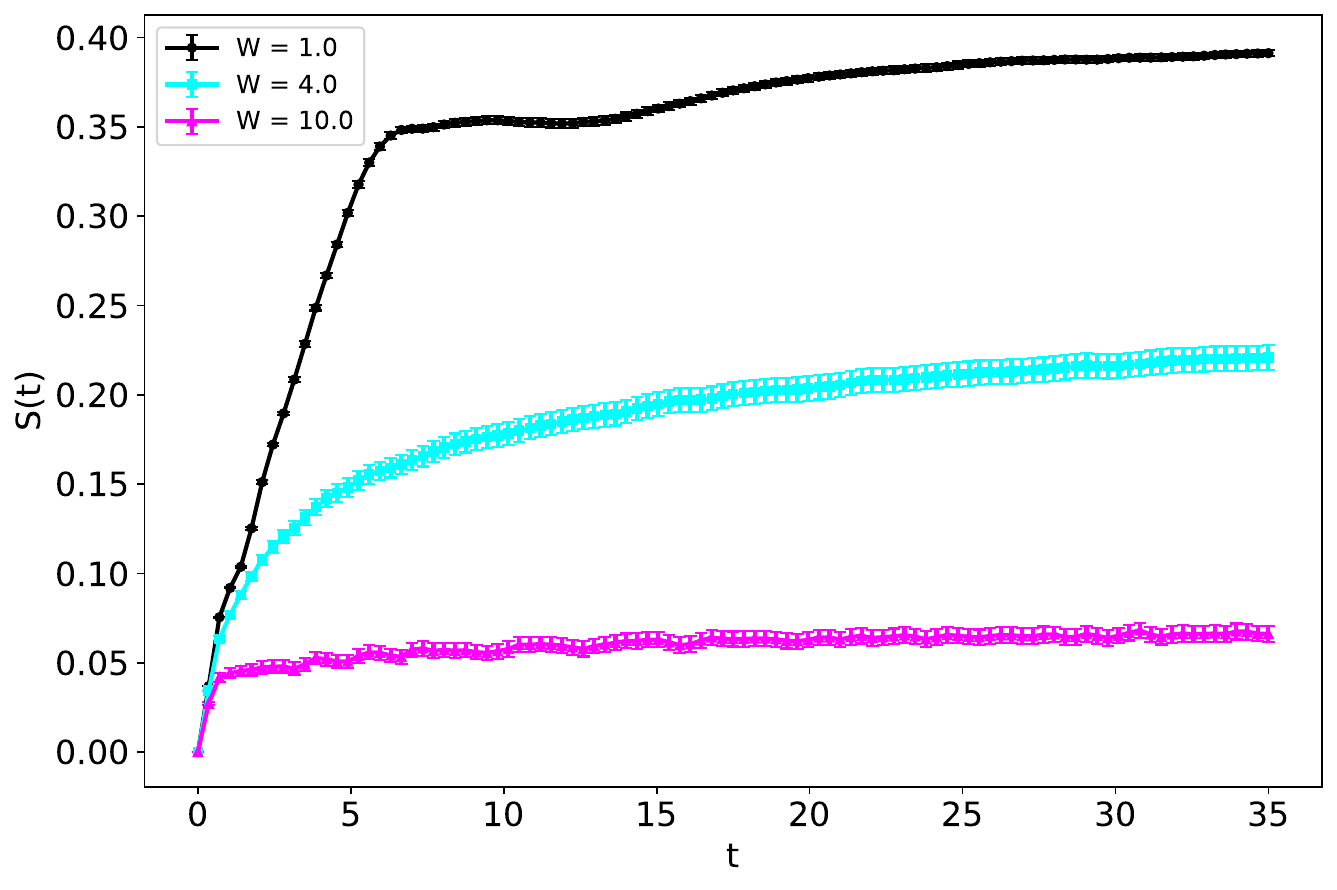}
		\caption{}
		\label{entanglement entropy disorder}
	\end{subfigure}

	\caption{(a) Entanglement entropy $S(t)$ as a function of time for different interaction strengths $U = 0$, $1$, $5$, $10$ at fixed disorder $W=10$ at $L = 12$. (b) Entanglement entropy $S(t)$ of $L = 10$ for different disorder strengths $W$. The transition from rapid growth (ergodic phase) to slow logarithmic growth (many-body localized phase) is clearly visible.}
	\label{fig:combined}
\end{figure}

The time evolution of the bipartite entanglement entropy $S(t)$ is shown in Fig. \ref{fig:combined} for varying the interaction strengths $U$ and disorder strengths $W$. The figure is presented as a two-panel plot, where Fig. \ref{Fig: entanglement entropy} illustrates the dependence on interaction strength at fixed strong disorder $W = 10$ and Fig.\ref{entanglement entropy disorder} shows the dependence on the disorder strength for a fixed interaction.

\begin{figure}[htbp]
	\centering
	\includegraphics[width=\linewidth]{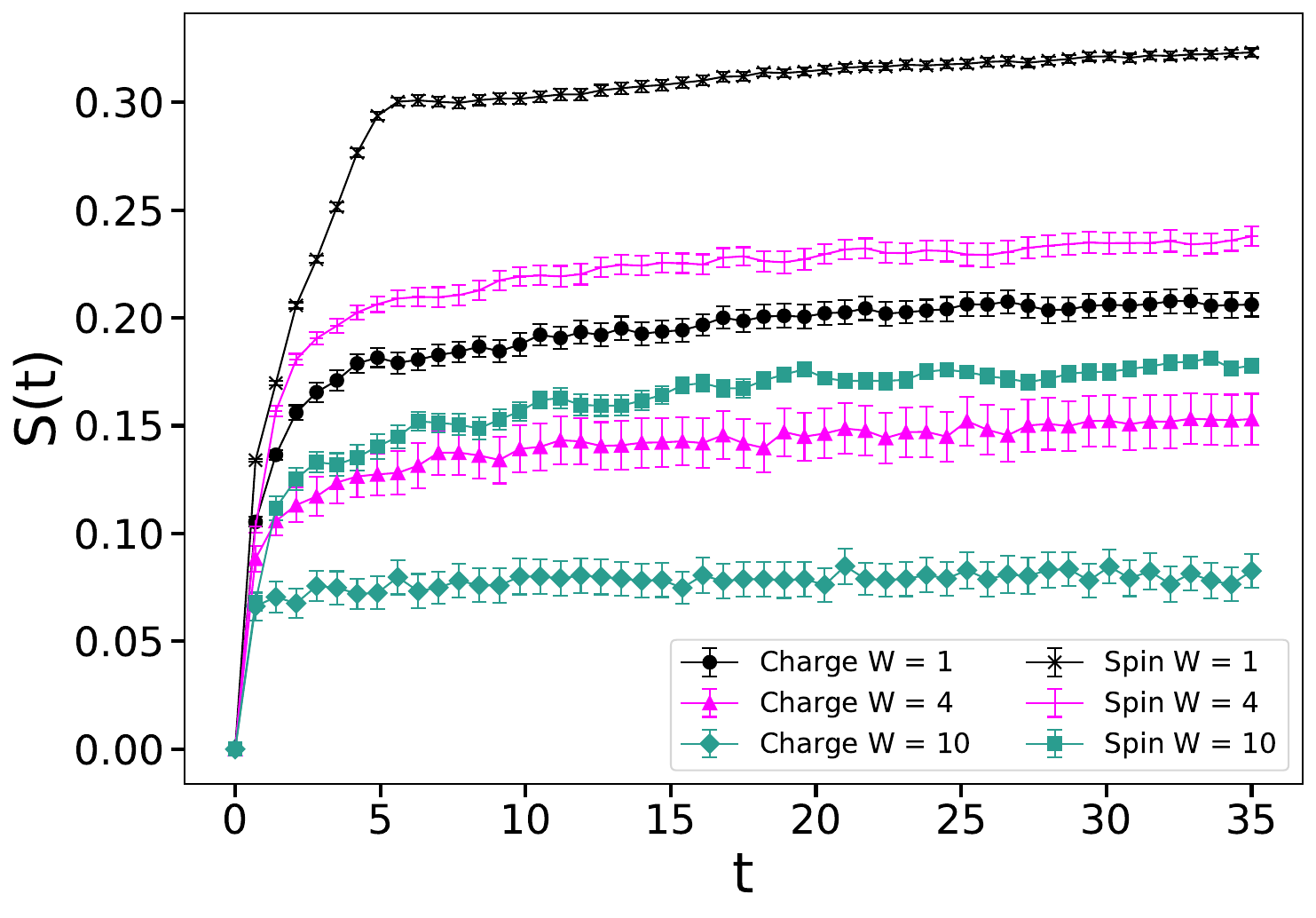}
	\caption{Time evolution of charge and spin entanglement entropy  $S_c(t)$ and $S_s(t)$ respectively for system sizes $L = 8$ , J = 1.0,U = 5.0 at disorder strength $W = 1.0, 4.0, 10.0$.}
	\label{fig:spin and charge entanglement dynamics}
	
\end{figure}

In Fig. \ref{Fig: entanglement entropy}, the non-interacting case $U = 0$ exhibits a rapid initial increase in entanglement entropy followed by an early-time saturation to a small value. This behavior is characteristic of Anderson localization, where the absence of interactions prevents the spread of entanglement beyond a finite length scale. Upon introducing finite interactions $U \gg 0$, the entropy displays a slow but persistent growth over time. The growth becomes more pronounced with increasing $U$, indicating that interactions enable the gradual buildup of quantum correlations through dephasing mechanisms. In particular, the time dependence of $S(t)$ at finite $U$ is consistent with a logarithmic growth law, $S(t) \sim \mathrm{log}\,t$, which is widely recognized as a hallmark of the many-body localized phase\cite{Serbyn20132,Bardarson2012, Yu2018}.
The effect of disorder strength is shown in Fig. \ref{entanglement entropy disorder}. For weak disorder $W= 1.0$, the entanglement entropy grows rapidly as linear with $t$ and saturates at a relatively high value, indicating ergodic dynamics and efficient thermalization. As the strength of the disorder increases $W = 4.0$, both the growth rate and the saturation value of $S(t)$ decrease, signaling the onset of localization effects. In the regime of strong disorder $W = 10.0$, the entropy exhibits very slow growth over time and does not reach saturation within the accessible time window. This slow dynamics is consistent with logarithmic entanglement growth and reflects the suppression of particle transport due to a strong disorder.

The entanglement dynamics displayed in Fig. \ref{fig:spin and charge entanglement dynamics}, further highlight the distinct behavior of spin and charge sectors. For weak disorder $W = 1.0$, the spin entanglement entropy grows rapidly during the initial stages of evolution, reflecting efficient spreading of spin correlations. In comparison, the charge entanglement entropy evolves more slowly, indicating that charge fluctuations are already substantially hindered by the disorder potential even at moderate timescales. With increasing disorder strength, the growth of the entanglement entropy is noticeably reduced, signaling suppression of transport and slower redistribution of quantum information across the lattice. In the strong disorder $W=10$, the entanglement growth becomes extremely slow and approach a weak logarithmic increase characteristic of localized interacting systems. The charge sector exhibits the strongest suppression, remaining close to a nearly saturated value over the accessible time window, whereas the spin sector retains comparatively larger growth, implying that spin excitations continue to evolve. The qualitative difference between the two sectors therefore reflects the dynamical separation of entangled spin and charge degrees of freedom in the presence of disorder and interactions.

\section{Conclusion}
\label{sec:conclusion}
We have investigated the non-equilibrium dynamics of the disordered Fermi-Hubbard model and identified a clear signature of partial many-body localization. By analyzing the time evolution of the sublattice imbalance, the spin-charge imbalance, and the entanglement entropy, we distinguish between ergodic and localized regimes. With increasing disorder strength, transport is suppressed, leading to the breakdown of thermalization and the persistence of memory of the initial state, as reflected in a finite long-time imbalance. The distinct relaxation behavior of spin and charge degrees of freedom highlights the role of interactions and provides evidence for partial decoupling between these sectors. Furthermore, the slow logarithmic growth of the entanglement entropy in the interacting regime confirms the presence of dephasing-driven dynamics typical of the localized phase. These findings provide a consistent picture of the interplay between disorder and interactions and contribute to the understanding of non-equilibrium dynamics in isolated systems described by the Fermi-Hubbard model. In this work, we observed that random disorder of comparable strength  drives the charge sector into a nonergodic regime. In contrast, the spin sector remains delocalized, as evidenced by the decay of spin correlations at long times. Consequently, spin-independent disorder fails to induce complete many-body localization.


\section*{Acknowledgements} 
V.A. gratefully acknowledges financial support from the Science and Engineering Research Board (SERB), under the Anusandhan–National Research Foundation (ANRF), Government of India, through the Core Research Grant (CRG/2023/001573). In addition, computational assistance from the PARAM Shavak (“Gryphon”) cluster, established in V.A.’s laboratory at NIT Jalandhar and developed by C-DAC  is appreciated.

\section*{DATA AVAILABILITY} 
The data that support the findings of this article are available upon reasonable request.

\end{document}